\documentclass{aa}
\usepackage{graphicx}
\usepackage{natbib}
\usepackage{amsmath}
\usepackage{amsfonts}
\bibpunct{(}{)}{;}{a}{}{,}

\newcommand{\tdelta}{\widetilde{\delta}}
\newcommand{\tnu}{\widetilde{\nu}}
\newcommand{\tkappa}{\widetilde{\kappa}}
\newcommand{\xb}{\boldsymbol{x}}

\newcommand{\lhbsa}{{\it LHBSa}}
\newcommand{\lhbsb}{{\it LHBSb}}
\newcommand{\va}{{\it VA}}

\begin{document}

\title{Some further comments on the paper \\ ``{\it biparametric scale adaptive filter design}''
for the detection of compact sources as proposed by M. L\'opez-Caniego et. al (2005, MNRAS 359, 993)}

   \author{R. Vio\inst{1}
          \and
          P. Andreani\inst{2}
          }

   \institute{Chip Computers Consulting s.r.l., Viale Don L.~Sturzo 82,
              S.Liberale di Marcon, 30020 Venice, Italy\\
              \email{robertovio@tin.it}
         \and
             INAF-Osservatorio Astronomico di Trieste
             via Tiepolo 11, 34131 Trieste, Italy \\
             European Southern Observatory
             Karl Schwarzschildstrasse 2, 85748 Garching, Germany\\
              \email{andreani@ts.astro.it}
             }

\date{Received .............; accepted ................}

\abstract{In this note we re-propose the arguments presented in \citet{vio05} examining 
the superiority of the {\it bi-parametric scale adaptive filter} (BSAF) when dealing with source detection
as claimed by \citet{lop05a} and confirmed in \citet{lop05b}. We suggest here that the dispute 
can be easily solved if these authors provide the community with a detailed derivation of a basic equation.
}
\titlerunning{Optimal Detection of Sources}
\authorrunning{R. Vio, \& P. Andreani}
\maketitle

\section{Introduction}

In recent times, a controversy has arisen concerning the real performance of the {\it bilinear scale adaptive
filter} (BSAF) introduced, in the context of one-dimensional signals, by \citet{lop05a} (hereafter \lhbsa)
for the detection of sources in a Gaussian background. In particular, \citet{vio05} (hereafter \va) raised three points:
\begin{itemize}
\item
a) in \lhbsa ~it is not clear how Eq.~(8) 
\begin{align} \label{eq:PH1a}
& p(\nu, \kappa| \nu_s) = \frac{\kappa}{\sqrt{2 \pi (1 -\rho^2)}} \nonumber \\
& \times \exp\left[ - \frac{(\nu - \nu_s)^2 + 
(\kappa-\kappa_s)^2 - 2 \rho 
(\nu-\nu_s) (\kappa-\kappa_s)}
{2(1-\rho^2)} \right],
\end{align}
with $\nu \in (-\infty, +\infty)$ and $\kappa \in (0, +\infty)$, was obtained \footnote{
We remind that $\nu$ and $\kappa$ are, respectively,
the normalized value and curvature of the observed signal in correspondence
to the true position of the source, whereas $\nu_s$ and $\kappa_s$ are the corresponding quantities of this.};
\item
b) no formal explanation is provided concerning the real usefulness of the filtering operation 
before the application of the detection test;
\item
 c) the numerical experiments presented in the paper are not sufficient 
to support the claimed {\it optimal} characteristics of BSAF in real applications.
\end{itemize}

It is our opinion that, even after the reply by \citet{lop05b} (hereafter {\lhbsb}), these issues have not 
yet been satisfactorily addressed. In the following we present some arguments to support our believes.

\section{The reply by L\`opez-Caniego et al. (2005) and our comments to it}

The objections to \va~ raised by \lhbsb~ are clearly illustrated in their Conclusions. For sake of clearness, we 
report them for a direct comparison with our comments:

\begin{itemize}
\item[a)] ``{\it \va~ have questioned an allegedly unproved formula in our work, which is in fact rigorously 
derived from previous works in literature (Rice 1954; Bardeen et al. 1986; Bond \& Efstathiou (1987).
Instead, \va~ proposed an incorrect formula.}'' -- We certainly believe that \lhbsa~ have rigorously 
derived Eq.~\eqref{eq:PH1a} from some already existing equations (in \va~ this point is never questioned).
However, also in \lhbsb, as well as in their original work, authors ``forget'' to provide the details.
We stress again
that the main problem is not the formula in {\it se} (which is certainly correct), but rather the conditions
of its validity that are not illustrated. From their text it appears that this equation was derived
with no restrictions concerning the relative position of the source with respect to the position of 
the peaks of the Gaussian background. But, in their numerical simulations the detection test is applied
only to the observed maxima, which -- after the filtering -- coincide with the true position 
of the sources. This implies that Eq.~\eqref{eq:PH1a} gives the probability density of maxima only
in presence of sources that satisfy some specific conditions.
This is confirmed by what authors write in Sect.~3 of 
\lhbsb: ``{\it We have derived the number densities of maxima in two cases: when a source is located at the position
of the maxima (not ``nearby the maxima'') and when there is no source}''. 
Now, a question arises: under which conditions
the peak of a source, superimposed to a background, is not moved to another position? Of course, one possibility is
when, in correspondence to the position of the true maximum of the source, the first derivative of the background
is equal to zero. This happens when
a source is constrained to be located in correspondence to a maximum of the underlying noise process. Actually, also
a minimum is possible, but in this case the additional condition $|\kappa_n| < \kappa_s$ has to be satisfied
in order that $\kappa$ does not become negative ($\kappa_n$ is the normalized curvature of the background at the true 
position of the source).
In practical applications where the approach is expected to provide an effective improvement
(i.e., sources with a smooth profile and white or quasi white-noise processes), this is a rare situation 
since $p(\nu_n, \kappa_n)$ peaks at $|\kappa_n| \gg
\kappa_s$. This is because, contrary to the sources, the noise processes are dominated by the contribution of the high
Fourier frequencies. Since the condition $\kappa_n = 0$ has to be strictly satisfied, the case of
a saddle point is even rarer if not an event with probability zero. In \lhbsa, but overall in \lhbsb 
\footnote{In Sect.~2, \lhbsb~ write ``{\it We would like to remark that Eq.~(3)}'' (here Eq.~\eqref{eq:PH1a}) ``{\it 
gives the number density of maxima coming from the combination of the background plus source.
This does not mean, at all, that the maximum of the source has to coincide
with the maximum of the noise process as stated by \va.}''}, authors seem to suggest 
that sources located at the position of the observed maxima can be obtained also with more general composition
of sources and background and no whatever explanation is provided here.
\hfill\break
We find peculiar that they did not consider as a necessary condition to settle this question
that is crucial in understanding the potentialities and the
limits of the approach. Indeed, our {\bf true criticism} concerns the lack of a detailed presentation of the
derivation of Eq.~\eqref{eq:PH1a}.

Another point to stress is that, contrary to what claimed in \lhbsb, in \va~ we do not propose any new formula
to substitute Eq.~\eqref{eq:PH1a}. In fact, the density function
\begin{align} \label{eq:PH1b}
& p(\nu, \kappa| \nu_s) = \frac{\kappa - \kappa_s}{\sqrt{2 \pi (1 -\rho^2)}} \nonumber \\
& \times \exp\left[ - \frac{(\nu - \nu_s)^2 + 
(\kappa-\kappa_s)^2 - 2 \rho 
(\nu-\nu_s) (\kappa-\kappa_s)}
{2(1-\rho^2)} \right],
\end{align}
with $\nu \in (-\infty, +\infty)$ and $\kappa \in (\kappa_s, +\infty)$, is {\bf explicitly indicated to correspond to an
unrealistic situation}. The point we raised is that, forcing the observed 
sources to have the peak
in correspondence of their true maximum, one favours the selection of the (filtered) sources that are located in 
correspondence to a maximum of the (filtered) noise, i.e., those whose likelihood $p(\nu, \kappa| \nu_s)$ is given by 
Eq.\eqref{eq:PH1b} (see below); \\
\item[b)] ``{\it In their second comment, \va~ criticize the lack of generality of the approach proposed in 
\citet{lop05a}. In particular, \va~ criticize the idea of filtering the data and applying the Neyman-Pearson detector
to the local maxima. Instead, they suggest that the generalization of the derivation of the Neyman-Pearson detector,
including not only amplitudes but also the second derivative of the field, should be done on purely
theoretical derivation, and the likelihood they propose is not general either, since it does not include
the first derivative of the field, that outside the maxima is not zero. Our approach, however, is consistent
and it leads to an improvement in the number of detections.}'' -- In practical
applications it is possible to work only with the observed maxima. In this context, the likelihood 
ratio $p_f(\tnu_n, \tdelta_n, \tkappa_n) / p_R(\nu_n, \kappa_n)$ indicated in \va~ is correct. In particular,
$p_f(\tnu_n, \tdelta_n, \tkappa_n)$ corresponds to the joint probability density function (PDF)
of the (normalized) height, the first and second derivatives of the background underlying the source in
correspondence to the {\it observed maximum}. This PDF is given, for example, in \citet{och90}.

Let us stress that the quantity $\tkappa_n$ is not necessarily equal to zero since there is no reason why
the observed maxima have to coincide with the  position of the true maximum of the source. 
In $p_R(.)$ it is $\delta_n=0$ since, under the hypothesis that no source is present in the observed signal,
the observed peaks correspond to the true peaks of the noise. We stress that the only reason why we have
introduced this equation is to provide a natural extension of the classic 
Neymann-Pearson technique to be used as benchmark. We are not interested to propose it as a possible
alternative to BSAF.

Concerning the {\bf effective usefulness in practical applications of the two steps (filtering + detection) procedure 
suggested by \lhbsa, we stress again that no rigorous theoretical arguments are presented}. The authors discuss
this aspect only in their Introduction but no mathematical formalization is given.

\item[c)] ``{\it In their third comment, \va~ criticize a set of numerical experiments designed to test our
theoretical arguments precisely for being designed to test our theoretical arguments. They suggest instead to make numerical
experiments in order to test what the theory does not say. We have derived the number density of maxima in two
cases: when a source is located at the position of the maxima (not ``nearby the maxima'') and where there is no
source. The way to test the hypothesis expressed by these formulas is exactly the one explained in L\'opez-Caniego
et al. (2005).}'' -- These sentences make evident what is the main limit of the proposed procedure: 
the theoretical arguments on which it is based are developed under the condition 
that an observed (filtered) source is located at the position of a maximum of the observed
(filtered) signal. {\bf Only under such condition it is correct what written in the Abstract} of \lhbsa:
``{\it The new filter 
includes as particular cases the standard matched filter and the scale-adaptive filter. Then by 
construction, the bi-parametric scale adaptive filter outperforms these filters}''. However, this
is not true in general situations.
If in the conclusion of \lhbsa~ this point had been clearly remarked, we should have no objections
to their work. The numerical experiments are used by authors not only
to test the correctness of their theory, but also to support
the use of the proposed approach in practical applications but such
experiments are not sufficient. The only thing they show is that {\bf BSAF overperforms the matched filter 
when the position of a (filtered) source coincides with the position of its observed maximum}.
But the occurrence of such a case is rather infrequent and in fact in the experiments by \lhbsa~ this particular
case happens only in the $10\%$ of the simulated
signals. So a natural question arises: what happens in more general situations? No answer to that is given.
It is not enough, as reported in \lhbsb, referring to the results by \citet{bar03}.
In that work the filters have no free parameters to be optimized. 

In \va~ we stressed one more point:
the investigation of the performance of the detection test only on the
sources that (after the filtering) show a peak placed at the position of their true maximum, is not safe. This is
because one favours the selection of the (filtered) sources that are located in correspondence 
to a maximum of the (filtered) noise, i.e., those whose likelihood $p(\nu, \kappa| \nu_s)$ is
given by Eq.\eqref{eq:PH1b}. It is not difficult to realize this in the discrete case (the case of 
interest for experimental
signals). For example, for a source with a Gaussian profile $G[i]$ located at the pixel $i_c$,
the addition of a white noise $w[i]$, with standard deviation $\sigma_w$, will not shift the position of the peak 
if $w[i_c] > w[i_c \pm 1]$. Working in the discrete domain (i.e., with a stairstep approximation
of the continuous functions) permits the alternative condition
\begin{equation} \label{eq:condition}
w[i_c \pm 1] - w[i_c] < \Delta,
\end{equation}
with $\Delta = G[i_c] - G[i_c \pm 1]$. 

In practical applications the detection tests are used in situations
of low signal to noise ratio (${\rm SNR}$) where $\Delta$ is a small quantity with respect to $\sigma_w$.
The consequence is that the condition~\eqref{eq:condition} is violated most of the times and the peak moves.
In order to verify this point, we have carried out a simple numerical experiment where a source with
Gaussian profile with dispersion set to three pixels and amplitude equal to one, has been superimposed to
a Gaussian white-noise background with standard deviation set to $0.35$. The level of noise has been 
chosen in such a way that, with a probability of false alarm is set to $10^{-3}$, in $1000$ replications
of the experiment the classic {\it matched filter} is able to detect all the sources. This is a case  
with good ${\rm SNR}$ and yet about $90\%$ of the cases where the position of the observed peak coincides with 
the position of the true maximum (approximately $33\%$ of the total number of simulations) is due to sources 
located at a maximum of the background, $0.6\%$ to sources located in correspondence to a minimum, and the rest 
to sources for which condition~\eqref{eq:condition} does not holds. As expected, the first percentage increases 
whereas the other two decrease when the ${\rm SNR}$ is worsen. This result supports the suspects raised in \va.
\end{itemize}

\section{Some additional comments to the questions raised by \citet{lop05b} to \va}

\begin{enumerate}
\item The reason why we intentionally included some basic elements about {\it matched filter} 
is only to unambiguously define the formalism as well as the theoretical framework in which developing our arguments.
 We firmly believe that this is a good practice, especially in the context of data analysis in Astronomy where {\it standard} formalism and notation
have not yet been fixed;
\item In \va, the likelihood ratio
\begin{equation} \label{eq:lr}
L(\xb,\nu, \kappa) = \frac{p(\xb, \nu, \kappa| H_1)}{p(\xb, \nu, \kappa| H_0)}
\end{equation}
is correctly written. As indicated by \lhbsb, the term $\nu$ is effectively redundant. We have maintained 
it to put in evidence the fact that 
the entire data set $\xb$ as well as the characteristic of the observed maximum are
considered. One could not like this choice, but it is allowed. The
likelihood ratio~\eqref{eq:lr} is not an alternative method to BSAF, but only an attempt 
to provide a theoretical justification of the \lhbsa~ approach. When the observed signal is filtered this 
implies the use of the entire data set $\xb$. After that, the detection test is based on the $\nu$ and $\kappa$ 
quantities. Since 
these two operations are optimized together, we guessed that the likelihood ratio~\eqref{eq:lr} could constitute
a theoretical explanation of the two-steps approach by \lhbsa. We are still convinced that this could be
a possible explanation. In any case, it is a minor point;
\item The complexity of BSAF in not a problem in {\it se}, rather with respect to the lack of flexibility
of the method. In fact, in practical applications it is not advisable to use a technique requiring a substantial modification
of the basic equations even when only a condition is relaxed; 
\item As last comment we would caution these authors to claim that the arguments and/or some equations in \va~ are wrong.
Indeed all of them are correct under the conditions that are explicitly given.
\end{enumerate}

\section{Conclusions}

In this work we have further discussed the arguments presented in \citet{vio05} concerning the
{\it bi-parametric scale adaptive filter} proposed by \citet{lop05a}. In particular, we still question the claimed
general optimal characteristics of this approach. Our claims are supported by the modality with which the
numerical experiments have been carried out by the authors.

A word of caution, the lack of rigorous theoretical foundation and clear assumptions made to
test a theory may lead to wrong conclusions, especially in a field such as cosmology where
all sky data are not very sensitive and the signal-to-noise ratio per pixel is at best equal to one.
Here we only urge the authors to prove the foundation of their work.
The simplest way would be a detailed presentation of the basic equation that is missing in the original work.

\end{document}